\input epsf
\input phyzzx
\hsize=40pc

\input BoxedEPS
\SetRokickiEPSFSpecial 
\HideDisplacementBoxes 

\catcode`\@=11 
\def\NEWrefmark#1{\step@ver{{\;#1}}}
\catcode`\@=12 

\def\square{\kern1pt\vbox{\hrule height 1.2pt\hbox{\vrule width 1.2pt\hskip 3pt
   \vbox{\vskip 6pt}\hskip 3pt\vrule width 0.6pt}\hrule height 0.6pt}\kern1pt}
\def\inbar{\,\vrule height 1.5ex width .4pt depth0pt}\def\IC{\relax
\hbox{$\inbar\kern-.3em{\rm C}$}}
\def\IH{\relax\hbox{$\inbar\kern-.3em{\rm H}$}}

\def\ket#1{| #1 \rangle}

\def\A{{\cal A}}
\def\B{{\cal B}}

\def\H{{\cal H}}
\def\F{{\cal F}}
\def\I{{\cal I}}

\def\K{{\cal K}}

\def\U{{\cal U}}

\def\p{\partial}

\def\B{{\cal B}}

\def\p{\partial}

\singlespace

\def\define#1#2\par{\def#1{\Ref#1{#2}\edef#1{\noexpand\refmark{#1}}}}
\def\con#1#2\noc{\let\?=\Ref\let\<=\refmark\let\Ref=\REFS
         \let\refmark=\undefined#1\let\Ref=\REFSCON#2
         \let\Ref=\?\let\refmark=\<\refsend}

\let\refmark=\NEWrefmark

\define\csft{B. Zwiebach, `Closed string field theory: Quantum action and the
Batalin-Vilkovisky master equation',
Nucl. Phys. {\bf B390} (1993) 33, hep-th/9206084.}

\define\cbi{A. Sen and B. Zwiebach, `Local background independence of
classical closed string field theory',
Nucl. Phys. {\bf B414} (1994) 649, hep-th/9307088.}

\define\qbi{A. Sen and B. Zwiebach, `Quantum background independence of closed
string field theory',
Nucl. Phys. {\bf B423} (1994) 580, hep-th/9311009.}

\define\coupling{A. Belopolsky and B. Zwiebach, `Who changes the string
coupling?',
Nucl. Phys. {\bf B472} (1996) 109, hep-th/9511077.}

\define\nonconf{B. Zwiebach, `Building string field theory around
non-conformal backgrounds',
Nucl. Phys. {\bf B480} (1996) 541, hep-th/9606153.}

\define\dilaton{O. Bergman and B. Zwiebach `The dilaton theorem and closed
string backgrounds',
Nucl. Phys. {\bf B441} (1995) 76, hep-th/9411047.}

\define\sabdilaton{S. Rahman and B. Zwiebach `Vacuum vertices and the
ghost-dilaton',
Nucl. Phys. {\bf B471} (1996) 233, hep-th/9507038.}

\define\fol{K. Ranganathan `A criterion for flatness in minimal area metrics
that define string diagrams',
Comm. Math. Phys. {\bf 146} (1992) 429.}

\define\bistruct{A. Sen and B. Zwiebach, `Background independent algebraic
structures in closed string field theory',
Comm. Math. Phys. {\bf 177} (1996) 305, hep-th/9408053.}

\define\openclose{B. Zwiebach, `Oriented open-closed string theory revisited',
MIT-CTP-2644, HUTP-97/A025, May 1997, hep-th/9705241.}

\define\ssft{A. Belopolsky, `New geometrical approach to superstrings',
IFP/103/UNC, March 1997, hep-th/9703183.}

\define\action{S. Rahman, `Geometrising the closed string field theory action',
MIT-CTP-2648, September 1997, hep-th/9709126.}

\define\recur{S. Rahman, `String vertices and inner derivations',
MIT-CTP-2645, June 1997, hep-th/9706128.}

\define\wittenone{E. Witten, `On background independent open-string field
theory',
Phys. Rev. {\bf D46} (1992) 5467, hep-th/9208027.}

\define\wittentwo{E. Witten, `Some computations in background independent
open-string field theory',
Phys. Rev. {\bf D47} (1992) 5467, hep-th/9210065.}

\define\samsonone{S. Shatashvili, `Comment on the background independent open
string theory',
Phys. Lett. {\bf B311} (1993) 83, hep-th/9303143.}

\define\wittenthree{E. Witten, `Quantum background independence in string
theory',
IASSNS-HEP-93/29, June 1993, hep-th/9306122.}

\define\samsontwo{S. Shatashvili, `On the problems with background
independence in string theory',
IASSNS-HEP-93/66, November 1993, hep-th/9311177.}

\define\sentwo{A. Sen, `On the background independence of string field theory',
Nucl. Phys. {\bf B345} (1990) 551.}

\define\senthree{A. Sen, `On the background independence of string field
theory. 2. Analysis of on-shell S-matrix elements',
Nucl. Phys. {\bf B347} (1990) 270.}

\define\senfour{A. Sen, `On the background independence of string field theory.
3. Explicit field redefinitions',
Nucl. Phys. {\bf B391} (1993) 550, hep-th/9201041.}

\singlespace
{}~ \hfill \vbox{\hbox{MIT-CTP-2649}
\hbox{
} }\break
\title{THE PATH TOWARDS MANIFEST BACKGROUND INDEPENDENCE}
\author{Sabbir A Rahman \foot{E-mail address: rahman@marie.mit.edu
\hfill\break Supported in part by D.O.E.
cooperative agreement DE-FC02-94ER40818.}}
\address{Center for Theoretical Physics,\break
Laboratory of Nuclear Science\break
and Department of Physics\break
Massachusetts Institute of Technology\break
Cambridge, Massachusetts 02139, U.S.A.}

\abstract{The set of string vertices is extended to include moduli spaces with
genus and numbers of ordinary and special punctures ranging over all integral
values $g, n, \bar n \geq 0$. It is argued that both the string background
and the B-V delta operator should be associated with the vertex
$\B^0_{0,1}$ corresponding to the once-punctured sphere. This leads naturally
to the proposal that the manifestly
background independent formulation of quantum closed string field theory is
given by the sum of the completed set of string vertices $\B = \sum_{g,n,\bar
n\geq0} \B^{\bar n}_{g,n}$, satisfying the classical master equation $\{ \B ,
\B \} = 0$.}
\endpage
\singlespace
\baselineskip=18pt

\chapter{\bf Introduction and Summary}

Since the proof by Sen and Zwiebach [\cbi,\qbi] of local conformal background
independence of both classical and quantum closed string field theory, which
stemmed from earlier work by Sen [\sentwo,\senthree,\senfour], much effort has
been exerted towards constructing a formulation manifestly independent of the
conformal background [\bistruct]. Similar efforts have also been made in the
framework of open string field theory by Witten and Shatashvili [\wittenone,
\wittentwo,\samsonone,\wittenthree,\samsontwo]. Recent work on
the dilaton theorem [\dilaton,\sabdilaton,\coupling] by Belopolsky, Bergman
and Zwiebach has shown that string backgrounds actually encode more data than
conformal backgrounds, namely the vacuum expectation value of the dilaton
field which governs the strength of the string coupling. This discovery
called into question the previously held assumption that string field theories
were built solely upon conformal field theories corresponding to classical
solutions. This led to a search for more general string backgrounds, towards
which strong progress has been made, notably in the recent work by Zwiebach
[\nonconf], where it was sketched how classical closed string field theory
might be built around non-conformal backgrounds without reference to any
conformal field theory. Background independence was not quite yet manifest, as
the string action was written as a function on the state space of an
underlying two-dimensional quantum field theory which represented the tangent
space to the space of backgrounds at that particular theory. However it did
suggest that a manifestly background independent formulation may be within
reach.

In two earlier papers [\recur,\action] by the author, the string field theory
operators $\p$, $\K$ and
$\I$ were expressed as inner derivations of the B-V algebra, and the recursion
relations for the string vertices were shown to take the form of a quantum B-V
master equation,
$$\half \{ \B , \B \} + \Delta \B = 0\,.\eqn\sabone$$
Here, $\B = \sum_{g,n,\bar n} \B^{\bar n}_{g,n}$ is the sum of the string
vertices for all non-negative integers $g,n,\bar n$ except $g=0, n+\bar
n\leq1$. The action was written in the simple form,
$$S = f(\B)\,.\eqn\sabonea$$
In this elegant geometrised form, the only background dependence is contained
in the string field $\ket{\Psi}$ and the fermionic state $\ket{F}$ required to
define the functional $f$ mapping the moduli spaces of decorated Riemann
surfaces to functions on the space of fields and antifields.

In the present paper, which is slightly more speculative than the earlier
works, We introduce the remaining negative-dimensional
moduli spaces, $\B^0_{0,0}$, $\B^0_{0,1}$ and $\B^1_{0,0}$ in order to
complete the set of string vertices, and incorporate them into the geometrised
formulation of the theory. We argue in \S 2.1 that $\B^0_{0,0}$ and
$\B^1_{0,0}$ decouple from the theory while the choice of string background
is encoded into the space $\B^0_{0,1}$. The latter does not decouple, and to
understand its antibracket action we propose in \S 2.2 a simple idea
about its sewing properties which would suggest
that the antibracket sewing of $\B^0_{0,1}$ should be identified with the
action of the B-V delta operator. The set of string vertices then becomes a
background independent algebraic structure containing complete information
about the theory. This leads us directly to postulate in \S 2.3 the manifestly
background formulation of quantum closed string field theory, where the
`geometrised' action is given by the sum of string vertices $\B =
\sum_{g,n,\bar n\geq0} \B^{\bar n}_{g,n}$, satisfying a classical master
equation, $\{ \B , \B \} = 0$.

\chapter{\bf Towards manifest background independence}

In the geometrised formulation of the theory described by
Eqns.\sabone-\sabonea\ the theory is encoded elegantly into the set of
string vertices satisfying a quantum master equation, as well as the string
field $\ket{\Psi}$ and the state $\ket{F}$ defining the mapping $f$ from
surfaces to functions. To construct this functional we need a knowledge of the
background which will determine the string field $\ket{\Psi}$ of ghost number
two, and the fermionic state $\ket{F}$ of ghost number three. The set of
string vertices are the background independent component of the formulation
while the string fields are the background-dependent component. If we are
somehow able to express this information in a background independent way, we
will have achieved our ambition of attaining a manifest background independent
formulation of the theory. A few suggestive observations will now
lead us to postulate the way in which we expect this goal to be realised.

\section{Completing the set of string vertices}

The object $\B$ is the formal sum of moduli spaces of decorated Riemann
surfaces $\B^{\bar n}_{g,n}$ for all positive values of $(g,n,\bar n)$ with the
exception of $\B^0_{0,0}$, $\B^1_{0,0}$ and $\B^0_{0,1}$. It would be
satisfying if there could be some way of incorporating them into the theory
and thereby completing the sum contained in $\B$.

Adding the spaces $\B^0_{0,0}$ and $\B^1_{0,0}$ should pose no problems. They
are represented by an unpunctured sphere and a sphere with a single special
puncture respectively, and their lack of ordinary
punctures implies that they do not
couple in any way to the remainder of the theory, contributing at most a
harmless constant to the action. The space $\B^0_{0,1}$ remains, and we shall
discuss this object now.

We recall from \S 4 of [\csft] that the algebraic structure
of the classical closed string field theory corresponds to a homotopy Lie
algebra $L_\infty$ defined by a set of multilinear graded commutative string
products $m_n = [ B_1 , \, . . . , B_n ]_0$. For a conformal background, it
was found that the product $m_0$ must vanish, $[ \, \cdot \, ]_0 = 0$, while
for string theory around a nonconformal background it was associated to some
non-vanishing ghost number three state $[ \, \cdot \, ]_0=\ket{\F}\in\H$.
Now, the
product $m_n$ is associated to the $n$-punctured vertex $\B^0_{0,n+1}$ with
fields inserted at $n$ of the punctures. So we expect the product $m_0$, which
determines the string background, to be associated with a once-punctured
sphere $\B^0_{0,1}$ with no insertions. Let us suppose that this is indeed
the case, so that the background independent object $\B^0_{0,1}$ actually
encodes the information about each particular string background.

Having introduced the spaces $\B^0_{0,0}$, $\B^0_{0,1}$ and $\B^1_{0,0}$, let
us also assume that the master equation for the $\B$-spaces is still satisfied.
We need to ask what the effect would be on the recursion relations. The spaces
$\B^0_{0,0}$ and $\B^1_{0,0}$ have no sewable punctures, and can be safely
ignored in this regard. However, $\B^0_{0,1}$ is not limited by this
restriction. Let us then define the operator $\U$ as follows,
$$\U\A = \{ \B^0_{0,1} , \A \}\,.\eqn\sabsixtytwo$$
Existence of $\B^0_{0,1}$ then implies that there would be a contribution of
$\U \B$ to our master equation. The vertex $\B^0_{0,1}$ has naive dimension
$-4$, whereas the boundary operator $\p$ is represented by the sewing of
$\B^0_{0,2}$ which is of dimension $-2$. The presence of $\U\B$ would therefore
signal the loss of the usual form of our usual recursion relations. If the
contribution from $\B^0_{0,1}$ is nilpotent, one may be able to recover
recursion
relations of some kind, but they will not agree with the geometrical identity
for the moduli spaces of Riemann surfaces. Furthermore we would also have
to explain the existence (or vanishing) of new objects $\{ \B^0_{0,1} ,
\B^0_{0,1} \}$, $\{ \B^0_{0,1} , \B^1_{0,1} \}$ and $\{ \B^0_{0,1} ,
\B^0_{0,2} \}$ which do not seem to have any interpretation in the theory as
yet.

With a view to resolving these issues, let us consider the surface which
$\B^0_{0,1}$ describes. It is a sphere with
a single coordinate disk whose boundary corresponds to a Hilbert space. This is
topologically equivalent to an infinite complex plane, the point at infinity
representing the puncture and the boundary being some contour around the
origin. This description is fine when there is no additional structure such as
endowed by a metric. Suppose now that we try to find a metric on the surface
which satisfies the minimal area conditions [\csft]. In the first place, we
note that attempting to put a metric on the sphere will always result in two
singular points. One of these may conveniently be associated with the puncture,
but the other one
has no such identification. Adding to this the requirement that nontrivial
closed loops have length $\geq 2\pi$, the geometrical description of
$\B^0_{0,1}$ is as an infinite cylinder, completely foliated by saturating
geodesics of length $2\pi$ [\fol], with a Hilbert space associated to the
puncture at one end [Fig. 1]. This description will now prove useful as
we attempt to identify the sewing of $\B^0_{0,1}$ with the B-V delta operation.

\midinsert
\epsfxsize 5in
\centerline{\epsffile{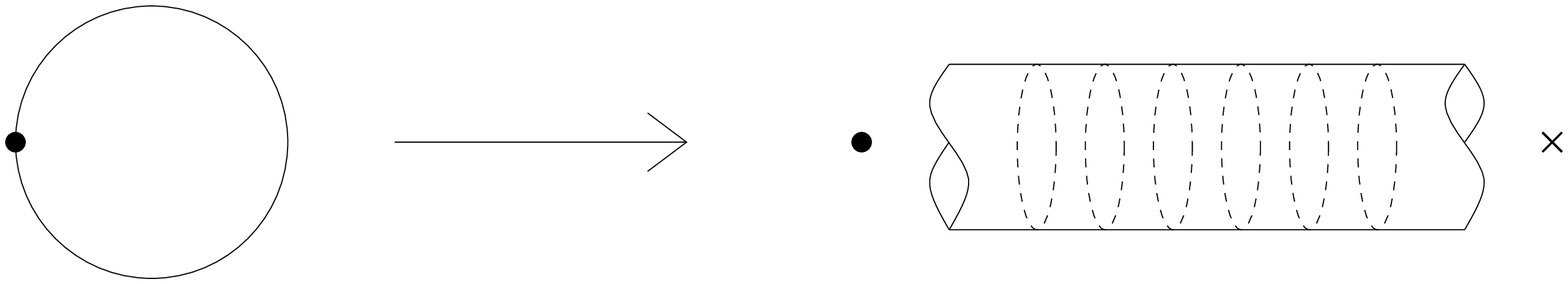}}
{Figure 1. Equipping the once-punctured sphere $\B^0_{0,1}$ with a
minimal area metric gives rise to an infinite cylinder with a Hilbert space
associated to the punctured end.}
\endinsert

\section{The sewing of $\B^0_{0,1}$ and the B-V delta operator}

Let us consider the effect of sewing $\B^0_{0,1}$ in view of the above
representation. Sewing it to another punctured surface will seemingly replace
a puncture on that surface with an infinite tube. This would be fine if the
tube ended at a puncture at the end of it with an associated Hilbert space, but
this is not the case here. We propose that in order for this sewing to be
well-defined, the boundless end of the cylinder must somehow lead to another
Hilbert space on sewing, the only Hilbert spaces available being those (should
they exist) which remain on the surface to which the space $\B^1_{0,1}$ is
being sewn [Fig. 2]. What this proposition would suggest is that sewing
$\B^1_{0,1}$ to a surface is equivalent to sewing together two punctures of
that surface! In other words, we argue that the operator we denoted as $\U$
earlier is none other than the B-V delta operator `$\Delta$'. As an immediate
corollary, the original quantum master equation for the complex $\B$ reduces
to the classical master equation, $\{ \B , \B \} = 0$.

\midinsert
\epsfxsize 6in
\centerline{\epsffile{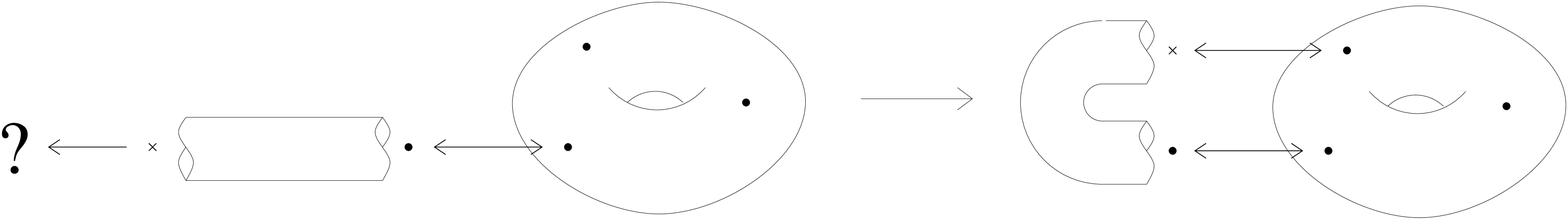}}
{Figure 2. Sewing $\B^0_{0,1}$ to a surface effectively sews
together two punctures of the surface.}
\endinsert

It was shown at the end of [\recur] that the B-V delta operator
cannot be an inner derivation of the B-V algebra, yet it may seem that we are
attempting to do precisely that here by identifying it with the sewing of
$\B^0_{0,1}$. However the sewing here does not coincide with antibracket sewing
as is clear from the description we have given above, and so
$\{ \B^0_{0,1} , \, \cdot \, \}$ is not an inner derivation on the algebra
despite appearances. That the identities Eqns.(2.37)-(2.40) of [\recur]
follow from an inner derivation
interpretation will be considered as merely a happy coincidence. One might
note that the objects $\{ \B^0_{0,1} , \B^0_{0,1} \}$
and $\{ \B^0_{0,1} , \B^1_{0,1} \}$ vanish from lack of sewable punctures,
while $\{ \B^0_{0,1} , \B^0_{0,2} \}$ gives the boundary of $\B^0_{1,0}$, as
discussed in [\action].

\section{Manifest background independent formulation}

If we suppose that the identifications we have suggested above are correct,
then it is no longer necessary to retain explicitly the action $S$, as all the
information required to encode the theory is already contained in the
background independent algebraic structure defined by the complete
set of string vertices. Indeed our final `manifest
background independent' formulation of the full quantum closed string field
theory would be given by the following `geometrical master action',
$$\B\,,\eqn\sabsixtynine$$
(where $\B = \sum_{g,n,\bar n\geq0} \B^{\bar n}_{g,n}$), satisfying the
`geometrical classical master equation',
$$\{ \B , \B \} = 0\,.\eqn\sabseventy$$
We shall conclude our analysis at this point.

\chapter{\bf Conclusion}

If the arguments which have been been presented in this chapter can be made
rigorous, we will have succeeded in finding a fundamental geometrical
representation of string field theory which underlies the usual formulation.
Furthermore, the unexpected appearance of the Batalin-Vilkovisky master
equation, which is satisfied by all gauge field theories strongly suggests
that this is a phenomenon which is not restricted to string theory, and may
be of much more general applicability, where our string vertices $\B$ would
be replaced by more general algebraic objects.

There are still many gaps in our understanding which would need to be
filled before we can be satisfied that we have found a manifestly background
independent formulation of string field theory. The arguments we have
brought forward merely suggest that our identifications are consistent, and
certainly do not represent proof of correctness. Besides this, our work is
largely dependent on the postulates of [\nonconf], some of which have yet to
be firmly established. As particular examples, we do not yet have explicit
expressions for the BRST operator $Q$, the fermionic state $\ket{F}$, the
one-form $F^{[1]}$ or the two-form $F^{[2]}$, and are therefore unable to
explicitly construct the action or extract physical states for general
backgrounds. Also, some of the issues which remained open at
the end of [\nonconf] still remain to be tackled. There are doubtless other
difficulties which we have not mentioned or of which we are as yet unaware.

Having said this, it is of interest to generalise our ideas to the case of
open-closed string theory [\openclose], and superstring
field theories as formulated in the language of [\ssft], in the hope of
eventually clarifying important issues such as the nature of the space of
string backgrounds and the existence of dualities. If such a program is
successful, one would expect that the results would also be of direct
relevance to field theory as a whole. Of course much more work needs to be
done before such ideas can be properly established but we hope nevertheless
that the ideas suggested in this paper will lead to some solid progress in our
basic understanding of string theory.

\ack
I would like to thank Barton Zwiebach and Robert Dickinson
for useful discussions, and am grateful to Chris Isham for allowing me
to complete some of this work at Imperial College, London.

\refout

\bye